\documentclass[pdftex,twocolumn,epjc3]{svjour3}          
\RequirePackage[T1]{fontenc}

\smartqed  

\RequirePackage{graphicx}
\RequirePackage{mathptmx}      
\RequirePackage{flushend}
\RequirePackage[numbers,sort&compress]{natbib}
\RequirePackage[colorlinks,citecolor=blue,urlcolor=blue,linkcolor=blue]{hyperref}

\journalname{Eur. Phys. J. C}
\usepackage{float}\usepackage{amsmath}
\usepackage[english]{babel}
\usepackage[utf8]{inputenc}
\usepackage[T1]{fontenc}
\usepackage{accents}
\usepackage{graphicx}
\usepackage{subfigure}
\usepackage{amssymb}
\usepackage{xcolor}
\usepackage[table,xcdraw]{}
\usepackage[normalem]{ulem}
\usepackage{xspace}
\usepackage{units}
\usepackage{multirow}
\usepackage{hyperref}

\newcommand{\NW}{$N_W$\xspace}

\newcommand{\Npart}{$N_{\text{part}}$\xspace}

\begin{document}

\title{How cross section fluctuations affect multiplicity and geometry in pA collisions}

\author{Chiara Le Roux\thanksref{e1,addr1}
}

\thankstext{e1}{e-mail: chiara.le\_roux@fysik.lu.se}

\institute{Dept. of Physics, Lund University, Sölvegatan 14A, S22362 Lund, Sweden\label{addr1}
}

\maketitle

\begin{abstract}

In this paper, a new Monte Carlo Glauber model is developed for pA collisions. It uses the hadronic cross sections calculated within the KMR model as implemented in the SHRiMPS minimum bias module of the SHERPA event generator. These cross sections are obtained as functions of impact parameter and, therefore, are ready for use in a Glauber model without additional assumptions regarding their impact parameter dependence. We compare the results obtained with those from a Black Disk model and from another model with colour fluctuations. It is shown that the KMR/SHRiMPS cross sections may present very good descriptions of the multiplicity distributions in pA collisions given that they show a long tail in the distribution of wounded nucleons. Moreover, it is shown that they also increase the anisotropy in the spacial distribution of wounded nucleons, which can be important in the description of the initial states of pA collisions. The generalization to A+A collisions is straightforward.

\end{abstract}

\section{Introduction}

The phenomenology of relativistic heavy ion collisions has been widely studied for decades now and the detection of a new state of matter made of quarks and gluons (QGP) has been well established \cite{Busza,BraunMunzinger2016,Shuryak2017,FOKA2016172}. The key instrument used in this discovery and in the study of the properties of the QGP is the comparison to pp collisions. However, a fair comparison can only be accomplished if one can precisely estimate, at least, the number of binary proton-proton interactions happening in heavy ion collisions. This way, one can properly scale the observables and, thus, make a fair comparison to pp collisions. But this number cannot be directly measured and, therefore, Glauber models arose as an important tool in making precise estimates \cite{dEnterria:2020dwq,Blaizot:2014qga,Broniowski:2007wp}.

However, Glauber modeling is not only important for experimental measurements. In order to connect theoretical predictions to experimental observations in collider experiments, one must make use of Monte Carlo event generators, which simulate the collisions of interest. There is a plethora of Monte Carlo models for the generation of heavy ion collision events \cite{URQMD,VENUS,HIJING,ANGANTYR}, all of which use Glauber models to predict the initial stages of collision. Different models use different techniques to go from number of participants to predicting actual experimental observables. In any case, it is well known that the charged particle multiplicity is strongly related to the number of participants \cite{miller2007glauber,Alver2008_PHOBOS_Glauber,dEnterria:2020dwq}. However, when applying such models to proton-nucleus collisions, there is a systematic underestimation of the small and large \Npart tails in the distribution \cite{ALICE:2025woy}. It has been previously shown that adding cross section fluctuations leads to an improvement in the description of the tails \cite{ALVIOLI2013347,Bierlich:2016smv}, nonetheless there is still a gap to be filled.

On the other hand, $N_{\text{part}}$ is not the only quantity that results from a Glauber model. As aforementioned, simulations of heavy ion collisions, for example using hydrodynamic models, use Glauber models to define the geometric configuration of the initial stages, e.g., how the energy is distributed in the transverse plane to the collision axis. Therefore, the interpretation of any anisotropies observed in final state distributions depend strongly on the Glauber model and it has been observed that fluctuations of the initial state are crucial to obtain such anisotropies \cite{Blaizot:2014qga,Broniowski:2007wp,Bozek:2013ska,Alver:2010gr}.

In this paper, we explore how different proxies to geometry and multiplicity are sensitive to different types of fluctuations. For that, we propose a new fluctuating Monte Carlo Glauber model based on the nucleon-nucleon cross sections calculated within the KMR model \cite{Ryskin:2009vq} as implemented in the SHRiMPS minimum bias event generator \cite{Akiba:2016llf}. The advantage of this model is that it contains a non-trivial impact parameter dependence that comes about from the same calculation that gives the integrated cross section, i.e., in a unified way. We compare this to the non fluctuating black disk model and to a black disk model that can account for integrated cross section fluctuations within the Glauber-Gribov colour fluctuation model \cite{PhysRevLett.67.2946} with a trivial impact parameter dependence. In section \ref{sec:shrimpsxs} we briefly summarize the KMR/ SHRiMPS model. After that, section \ref{sec:newMCG} presents the implementation of these and the other models' cross sections in the new Glauber model and, finally, section \ref{sec:results} presents and discusses the results followed by the conclusion in section \ref{sec:conclusions}.

\section{KMR/SHRiMPS cross sections}
\label{sec:shrimpsxs}

The cross sections implemented in the new Glauber Model are based on the work by KMR \cite{Ryskin:2009vq} as implemented in the SHRiMPS \cite{Akiba:2016llf} module of the SHERPA event generator \cite{Sherpa:2019gpd}. This model allows for the calculation of soft QCD observables in high energy hadronic collisions. Since such processes live in the realm of non perturbative physics, standard perturbation theory cannot help. In that case, it is worth resorting to phenomenological models that predate QCD and had great success at describing soft observables. In the low energy limit, Regge theory provides one such method. This theory models hadron interactions as exchanges of particles called reggeons\footnote{Recall that Regge theory predates the parton model.}, which are identified as the poles of the scattering amplitudes continued to the complex plane \cite{Barone:2002hpd}. Regge theory has also been extended for high energy hadron collisions although, in that case, it requires the introduction of a reggeon called \textit{pomeron}, in order to describe the experimentally observed result that the total cross section increases with energy. After the advent of QCD, the pomeron was reinterpreted as an exchange of two or more gluons, thus making contact with the current understanding of QCD. The KMR model follows this same line by modeling hadron interactions as exchanges of relativistic pomerons, including pomeron rescatterings, and extending the partonic interpretation as a means to calculate cross sections. Bellow, we review the main points addressed by KMR as implemented in the SHRiMPS model.

In quantum mechanics, the whole dynamics of a scattering process is contained in the scattering matrix, $S$. By decomposing the incoming and outgoing particles in (orthogonal) partial waves of angular momentum $l$, it can be shown that the elastic scattering matrix becomes $S_{el}=e^{2i\delta_l}$, where $\delta_l$ is the phase shift to each of the partial wave amplitudes resulting from the scattering. One can separate the trivial part of the scattering matrix from the transition matrix $T_{el}$ by writing $S_{el}=1+iT_{el}$. Now, since angular momentum is conserved and the partial waves of different $l$ do not mix, this means that the impact parameter $b=2l/\sqrt{s}$ will be fixed throughout the interaction and $S_{el}$ (and conversely $T_{el}$) can be cast as functions of $b$ and of the squared center of mass energy $s$: $S_{el}(s,b)=1+iT_{el}(s,b)$. In this case, the transition matrix carries the following property:

\begin{equation}
    \text{Im}T_{el}(s,b) = 1-e^{-\Omega(s,b)/2},
\end{equation}

\noindent where $\Omega(s,b)$ is related to the phase shift $\delta_l$ and is known as eikonal or opacity. It is the result of the Fourier transformation of the momentum space amplitudes into impact parameter space. Therefore, to compute $\Omega(s,b)$, one must draw all the possible reggeon exchange diagrams. With that in hands, and making use of the optical theorem, which relates the total cross section to the forward scattering amplitude, the impact parameter dependent cross sections can be calculated as:

\begin{align}
    \label{eq:singeikxs}
    \frac{d\sigma_{tot}}{d^2b}(s,b) &= 2 T_{el}(s,b) = 2(1-e^{-\Omega(s,b)/2}),\nonumber\\
    \frac{d\sigma_{el}}{d^2b}(s,b) &= |T_{el}(s,b)|^2 = (1-e^{-\Omega(s,b)/2})^2,\nonumber\\
    \frac{d\sigma_{inel}}{d^2b}(s,b) &= 2 T_{el}(s,b) - |T_{el}(s,b)|^2 = (1-e^{-\Omega(s,b)}),
\end{align}

\noindent where we have used the eikonal approximation, that is, at high energies, $\text{Re}T_{el}/\text{Im}T_{el}<<1$ and, thus, the real part of $T_{el}$ can be ignored. This formalism, however, cannot account for low mass diffraction, which is a consequence of the internal structure of hadrons. Therefore, to do that, one can describe hadrons as superpositions of Good-Walker (GW) states \cite{Good:1960ba}. They are the states that diagonalize the transition matrix $T_{el}$ and, therefore, $\langle\phi_i|T_{el}|\phi_j\rangle=\delta_{ij}$. But if a hadron is in a superposition of such states, $|h_i\rangle=\sum_ja_j|\phi_j\rangle$, then it can be inelastically diffracted, meaning that, after the scattering, it will be in a different superposition of GW states, $|h_f\rangle=\sum_jb_j|\phi_j\rangle$. In this case,  the elastic scattering amplitude is: 

\begin{align}
    \label{eq:avgT}
    T_{el}&=\langle h_i|T_{el}|h_i \rangle 
          = \sum_j |a_j|^2 \langle \phi_j|T_{el}|\phi_j\rangle =\nonumber \\
          &= \sum_j |a_j|^2 (T_{el})_j 
          =\langle T_{el}\rangle_{h_i},
\end{align}

\noindent i.e., it is given by the average of the components of $T_{el}$ over the initial superposition of GW states. Similar expressions to those in Eqs. \ref{eq:singeikxs} can then be found and, in this case, a diffractive cross section emerges in addition to the elastic and inelastic (non-diffractive) ones.

Now, if we are treating the case of hadron-hadron collisions, we should treat both the projectile and target as superpositions of GW states. This means that Eq. \ref{eq:avgT} will contain two averages, one over projectile, $|p\rangle=\sum_ia_i|\phi_i\rangle$, and one over target, $|t\rangle=\sum_jb_j|\phi_j\rangle$ states:

\begin{align}
    \label{eq:avgTpt}
    T_{el}&=\langle p|\langle t|T_{el}|t \rangle|p \rangle \nonumber \\
          &= \sum_j |a_j|^2\sum_i |b_i|^2 \langle \phi_j|\langle \phi_i|T_{el}|\phi_i\rangle|\phi_j\rangle \nonumber \\
          &= \sum_{i,j} |a_j|^2|b_i|^2 (T_{el})_{ij} 
          =\langle T\rangle_{p,t}.
\end{align}

This way, one can account for the possibility of each (or both) of the incoming hadrons diffracting from one superposition of GW states in the ground state (e.g., the proton) to another superposition of GW states forming an excited state (e.g., N(1440)). This means that we have now multiple channels for inelastic interactions and, therefore, call this the multi channel eikonal formalism. In this case, $T_{el}(s,b)$ becomes a matrix of dimension $N_{GW} \times N_{GW}$ (where $N_{GW}$ is the number of GW states), and, therefore, so is the eikonal $\Omega_{ij}(s,b)$. Analogous expressions to those in Eqs. \ref{eq:singeikxs} can now be obtained, including those for single and double diffraction (when one or both of the hadrons go to the excited state respectively). The challenge, however, remains on how to compute the eikonals in order to obtain the cross sections as functions of the impact parameter.

As a first approximation, one can simply consider that the eikonals are such that:

\begin{equation*}
    \text{Im}T_{el}(s,b) = 
    \begin{cases}
        1 &, b \leq R \\
        0 &, b > R.
    \end{cases}
\end{equation*}

This corresponds to a Black Disk approximation and results in $\sigma_{tot}=2\pi R^2=\sigma_{el}/2=\sigma_{inel}/2$, i.e., no diffraction can take place in this case. Other phenomenological assumptions can be made about the impact parameter dependence of the amplitudes and, therefore, of the cross sections, and the parameters of the model can be fit so as to reproduce experimentally measured integrated cross sections or $d\sigma/dt$ for example\footnote{Notice that it is not possible to directly measure impact parameter dependent cross sections since the impact parameter is not accessible to experiments.}. It is even possible to include diffraction by allowing for the amplitude to fluctuate with certain probabilities since the diffractive cross sections are directly related to fluctuations of $T_{el}$, namely, for example, $d\sigma_{SD}/d^2b = \langle T^2\rangle - \langle T \rangle^2$, for the single diffraction case. Nonetheless, in the KMR model, the eikonals are computed from a set of evolution equations, thus allowing for the calculation of the $b$-dependent differential cross sections.

To understand how this is done, it is worthwhile noticing that, in the multichannel eikonal formalism, the inelastic non-diffractive cross section is equal to:

\begin{align}
    \frac{d\sigma_{inel}}{d^2b} &= \langle 2T_{el}\rangle_{p,t} - \langle T_{el}^2\rangle_{p,t} \nonumber \\
    &=\sum_{i,j}^{N_{GW}}|a_i|^2|a_j|^2 (1-e^{-\Omega_{ij}}) \nonumber \\
    &=\sum_{i,j}^{N_{GW}}\frac{d\sigma_{inel}^{(ij)}}{d^2b},
\end{align}

\noindent where $\Omega_{ij}\ge0$. Therefore, for a given channel $(ij)$, the inelastic cross section is equal to $d\sigma_{inel}^{(ij)}/d^2b=(1-e^{\Omega_{ij}}) \le 1$, which not only allows for the interpretation of the differential cross section as a probability, but also shows that $e^{-\Omega_{ij}}$ gives the probability that no inelastic interaction happens via the channel $(ij)$.

In the KMR model, each eikonal $\Omega_{ij}$ is interpreted as the flux of color being exchanged by the incoming hadrons\footnote{I.e., a cut reggeon in the language of Regge theory.} and it is given by the convolution of the parton density of the GW state $i$ in the presence of $j$, $\Omega_{i(j)}$ and that of the GW state $j$ in the presence of $i$, $\Omega_{(i)j}$. Notice that the dependence on the squared center of mass energy $s$ can be written as a dependence on rapidity instead by using $Y=\ln(\sqrt{s}/m_h)$, where $m_h$ is the hadron mass. Therefore:

\begin{align}
    \label{eq:eikonal}
    \Omega_{ij}(Y,B)=\frac{1}{\beta_0^2}\int d^2b_1 d^2b_2 &\delta^2(\mathbf{B}-\mathbf{b}_1-\mathbf{b}_2) \cdot \nonumber \\
    &\cdot\Omega_{i(j)}(y,b_1) \Omega_{(i)j}(y,b_2),
\end{align}

\noindent where $Y$ is the beam rapidity; $B$ is the impact parameter between the incoming hadrons; $b_i$ are the parton positions relative to the respective hadrons in the transverse plane and $\beta_0^2$ is a normalization factor. The reason why the parton densities of each GW state depend on the other GW state's is because they can be affected by the rescatterings between the partons and, since these scattering can happen anywhere in rapidity, the eikonal will be independent of $y \in [-Y/2,Y/2]$.

Therefore, to calculate the eikonal $\Omega_{ij}$, one needs the parton densities. As a first approximation it can be reasonable to assume that $d\Omega_{i(j)}/dy \propto \Delta \Omega_{i(j)}$ and $d\Omega_{(i)j}/dy \propto \Delta \Omega_{(i)j}$. That is, $\Delta$ is a probability that a parton will emit additional partons and, therefore, the rate of increase of the parton density increases. This way, the two parton densities would be independent of each other and a simple solution for each of them can be found. However, one must also account for the fact that, apart from emitting other partons, they can also be absorbed by the presence of the incoming parton density. This can be taken care of by introducing another factor to the evolution equations, which becomes:

\begin{align}
    \label{eq:eikevolution}
    \frac{d\Omega_{i(j)}(b_1,b_2)}{dy} &=+\mathcal{W}_{(ij)}(b_1,b_2)\cdot\Delta \cdot\Omega_{i(j)}(b_1,b_2) \nonumber \\
    \frac{d\Omega_{(i)j}(b_1,b_2)}{dy} &=-\mathcal{W}_{(ij)}(b_1,b_2)\cdot\Delta \cdot\Omega_{(i)j}(b_1,b_2),
\end{align}

\noindent where $\mathcal{W}_{(ij)}$ parametrizes the absorptive correction. In the SHRiMPS \cite{Akiba:2016llf} implementation of the model, this factor is written as:

\begin{align}
    \mathcal{W}^{(ij)}&(b_1,b_2)
    = \nonumber \\
    &=\left[ \frac
        {1-e^{\left[-\frac{\lambda}{2}
            \Omega_{i(j)}(b_1,b_2)\right]}}
        {\frac{\lambda}{2}\Omega_{i(j))}(b_1,b_2)}
        \right]
        \left[\frac
        {1-e^{\left[-\frac{\lambda}{2}
            \Omega_{(i)j}(b_1,b_2)\right]}}
        {\frac{\lambda}{2}\Omega_{(i)j}(b_1,b_2)}
        \right].
\end{align}

The parameters $\Delta$ and $\lambda$ are constants. $\Delta$, which iterates the emission of partons by the initial parton density, is given by the pomeron intercept $\Delta=\alpha_\textbf{P}-1\approx0.3$, whereas $\lambda$ drives the recombination of those partons, thus being given by the triple pomeron vertex from the BFKL formalism, $\lambda \approx 0.25$. With this, the only ingredient missing in order to solve the evolution equations \ref{eq:eikevolution} are the boundary conditions. In the SHRiMPS model they are given by modified dipole-like form factors:

\begin{equation}
    \label{eq:formfactors}
    \mathcal{F}_{1,2}(q_\perp) = \beta_0^2(1 \pm \kappa)
    \frac
    {e^{-\frac{\xi(1\pm\kappa)q_\perp^2}{\Lambda^2}}}
    {(1+\frac{(1\pm\kappa)q_\perp^2}{\Lambda^2})^2},
\end{equation}

\noindent Fourier transformed into impact parameter space:

\begin{align}
    \Omega_{i(j)}(-Y/2,b_1)&=F_i(b_1) \nonumber \\
    \Omega_{(i)j}(-Y/2,b_2)&=F_j(b_2).
\end{align}

With this, the evolution equations for the eikonals can be solved and the cross sections can be calculated. The free parameters of the model are: $\beta_0^2$, showing in Eqs. \ref{eq:eikonal} and \ref{eq:formfactors}, a quantity with dimensions of area (or cross section) that is of the order of the total hadronic cross section; the $\Lambda^2$ parameter, related to the form factors' fall off in momentum space; and the dimensionless parameters also showing in the form factors: $\kappa$ and $\xi$. These parameters were tuned to experimental measurements and, in section \ref{sec:newMCG}, two of those tunes will be discussed. The SHRiMPS model also uses two GW states to decompose the proton and its excited state N$^*$(1440), which means that the diffractive cross section describes low mass diffraction, whereas high mass diffraction is included in the inelastic cross section. For both of the states, the probabilities for each GW state are 0.5 in even (proton) or odd (N$^*$(1440)) combinations. More details on the KMR model can be found in Ref. \cite{Ryskin:2009vq} and references therein and, on the SHRiMPS implementation, in section 2.7 of Ref. \cite{Akiba:2016llf}.
    
\section{The new Monte Carlo Glauber model}
\label{sec:newMCG}

The new Glauber model was built in the usual way: 

\begin{enumerate}
    \item Distribute nucleons inside the nucleus according to the Woods-Saxon distribution with a minimum separation of \unit[0.8]{fm} between nucleons;
    \item Select an impact parameter for the proton-nucleus system according to the probability $P(b) \propto bdb$;
    \item Interpret the impact parameter dependent pp cross section for inelastic scattering in each given model as the probability for interaction between the projectile proton and each target nucleon.
\end{enumerate}

At each event, one can count the number of inelastically wounded nucleons\footnote{In the context of this paper, "wounded nucleon" or \NW stands for a nucleon that underwent an inelastic non-diffractive interaction.} (\NW) in a given pA collision and also obtain their geometric distribution.

In order to explore the effect of fluctuations on the results of the Glauber model, we use three different models for the pp cross sections:

\subsection*{Black Disk (BD) model}

This the simplest of the models used here and does not include any fluctuations. The impact parameter dependent cross section in this case is given by:

\begin{equation}
    \frac{d\sigma^{BD}}{d^2b} = \Theta(R_{BD} - b),
\end{equation}

\noindent where $R_{BD}$ is chosen such that the total inelastic cross section matches the integrated inelastic cross section \textit{averaged} in projectile and target states calculated within the SHRiMPS model.

\subsection*{Glauber-Gribov (GG) model}

This Glauber-Gribov inspired model is described and implemented in the PHOBOS Monte Carlo model \cite{Alver2008_PHOBOS_Glauber}. It is based on the idea that the nucleons can fluctuate \cite{Good:1960ba}, which leads to cross section fluctuations. It is a black disk model with integrated cross section (and, therefore, black disk radii) fluctuations. The cross section for inelastic scattering is sampled from:

\begin{equation}
    \label{eq:ggprob}
    f(x) = \frac{\rho}{\lambda\sigma_0}\frac{x}{x + \lambda}e^{-\left(\frac{x-\lambda}{\lambda\Omega}\right)^2},\quad x=\frac{\sigma_{inel}}{\sigma_0},
\end{equation}

\noindent where $\rho$ is a normalization factor, $\Omega = 1.01$\footnote{This is one of the default values used, e.g., in \cite{Adam:2015nra}.}, $\lambda=\langle\sigma_{inel}\rangle/\langle\sigma_{tot}\rangle$, and $\sigma_0$ is set such that $\langle \sigma_{inel}\rangle=\sigma_0\langle x\rangle=\sigma_0\int x f(x) dx$ gives the average inelastic cross section calculated within the SHRiMPS model. As will be discussed in the next subsection, two different tunes were used for the SHRiMPS model. For tune 1, we obtain $\langle\sigma_{tot}\rangle=$\unit[9.1]{fm$^2$} and $\langle\sigma_{inel}\rangle=$\unit[6.3]{fm$^2$} (i.e., $\lambda=0.69$), achieved by setting $\sigma_0=$\unit[6.95]{fm$^2$}, whereas, for tune 2, $\langle\sigma_{tot}\rangle=$\unit[8.9]{fm$^2$} and $\langle\sigma_{inel}\rangle=$\unit[6.3]{fm$^2$} (i.e., $\lambda=0.71$), and, thus, $\sigma_0=$\unit[6.8]{fm$^2$}, therefore, both tunes give very similar values.

\subsection*{KMR/SHRiMPS model}

The KMR/SHRiMPS (SH) cross sections are described in section \ref{sec:shrimpsxs}. They can account for fluctuations of the integrated cross section by allowing both target and projectile to fluctuate between two possible Good-Walker states with equal probabilities. In the Glauber model, this is done by randomly assigning one of two states with probability 0.5 to each of the target and projectile nucleons event-by-event. But, most importantly, the impact parameter dependence, in this case, results from the calculation described in section \ref{sec:shrimpsxs}, it is not imposed. As mentioned before, the free parameters of the model are fixed by tuning to experimental measurements. Table \ref{tab:shrimpspar} presents the values for each of the parameters used with each tune.

\begin{table}[htbp]
    \centering
    \caption{Two sets of parameters used in the calculation of the KMR/SHRiMPS cross sections.}
    \label{tab:shrimpspar}
    \begin{tabular}{|c|c|c|}
        \hline
        Parameter & \textbf{tune 1} & tune 2 \\
        \hline
        $\beta_0^2$   & \textbf{\unit[$\textbf{5.397}$]{mb}}   & \unit[$3.875$]{mb}   \\
        $\kappa$   & \textbf{0.5966}   & 0.4583   \\
        $\xi$  & \textbf{0.1014}   & 0.1992   \\
        $\Lambda^2$  & \textbf{\unit[\textbf{1.683}]{GeV$^\textbf{2}$}}   & 1.205 GeV$^2$   \\
        \hline
    \end{tabular}
\end{table}

\begin{figure}[!ht]
    \centering
    \includegraphics[width=0.63\linewidth]{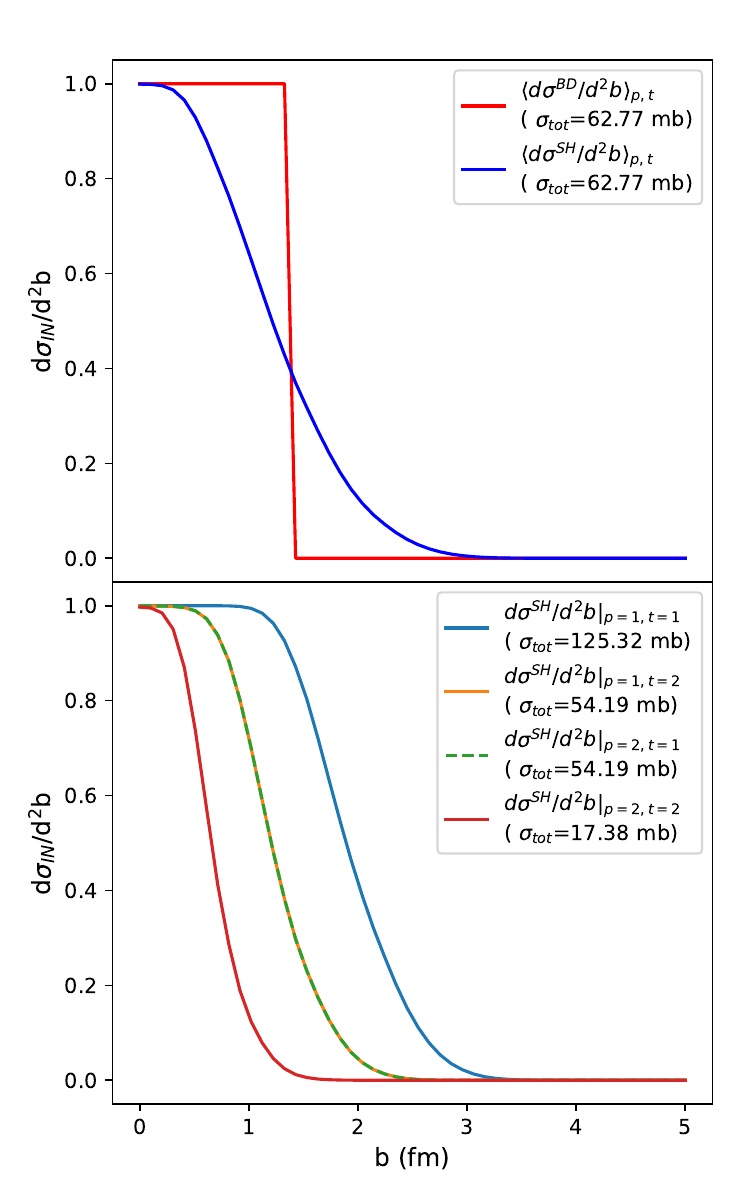}
    \caption{Impact parameter dependence of the inelastic cross section within the BD and SH models at $\sqrt{s}=\unit[5020]{GeV}$ with parameter settings from tune 1 (Tab. \ref{tab:shrimpspar}). Top: BD cross section and SH cross section averaged over target and projectile states; Bottom: SH cross sections for different combinations of projectile and target GW states.}
    \label{fig:xs-vs-b-tune1}
\end{figure}

Both tunes yield very similar results for the integrated cross sections (as aforementioned) and their impact parameter profiles are also approximately the same. That shows that the model is rather stable and, therefore, tune 1 was chosen for the purposes of the results shown in the next section. With this choice, the cross sections as functions of the impact parameter for the BD model and for the different combinations of Good-Walker states in the SH model at a center of mass energy of $\sqrt{s}=\unit[5020]{GeV}$ is shown in Fig.~\ref{fig:xs-vs-b-tune1}.

\begin{figure}[!ht]
    \centering

    \includegraphics[width=1.0\linewidth]{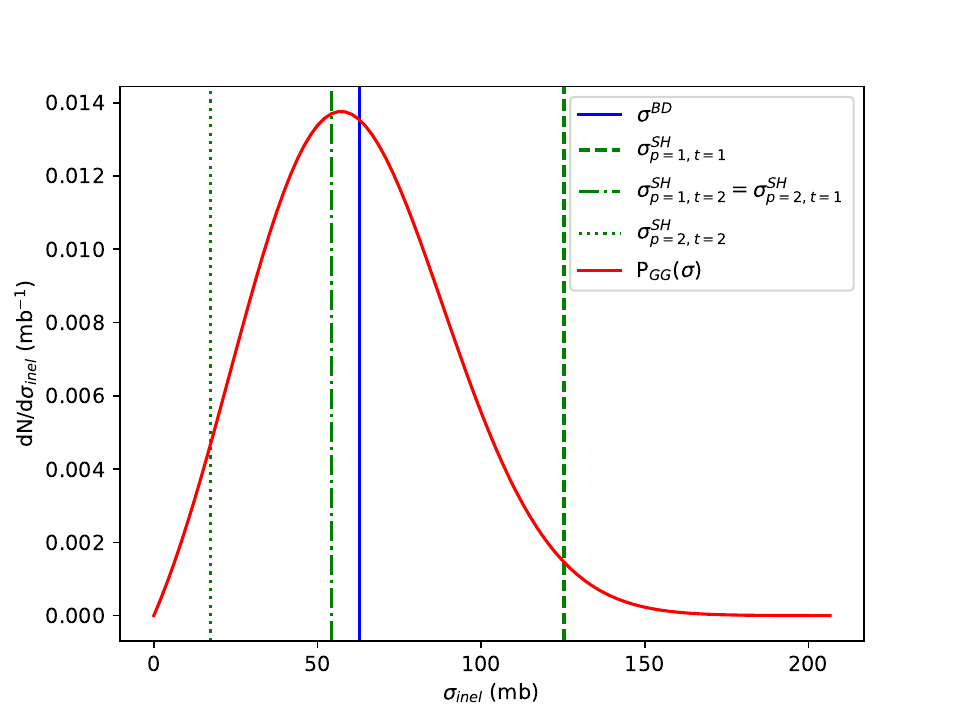}
    
    \caption{All possible values of the integrated inelastic cross section within the SH model alongside the value of the integrated inelastic cross section in the BD model and the probability distribution for the integrated inelastic cross section within the GG model.}
    \label{fig:totalxss}
\end{figure}

The GG model has the same impact parameter dependence as the BD but with a fluctuating radius. Fig.~\ref{fig:totalxss} shows the probability distribution function for the integrated inelastic non-diffractive cross section within the GG model along with the values for the same integrated cross sections with the BD and SH models.

Therefore, we see that the BD cross section does not fluctuate, whereas the GG cross section fluctuates continuously according to Eq. \ref{eq:ggprob} and reaches more extreme values than the ones obtained in the SH model. On the other hand, the SH cross section fluctuates between 3 different values depending on the combination of GW states of the colliding nucleons, with the intermediate value being twice as likely as each of the other ones since it is obtained with two different configurations of GW states. In the next section, we explore the consequence of the different characteristics of each model on the multiplicities and geometries of pA collision events.

\section{Results}
\label{sec:results}

\begin{figure}[!ht]
    \centering
    \includegraphics[width=0.8\linewidth]{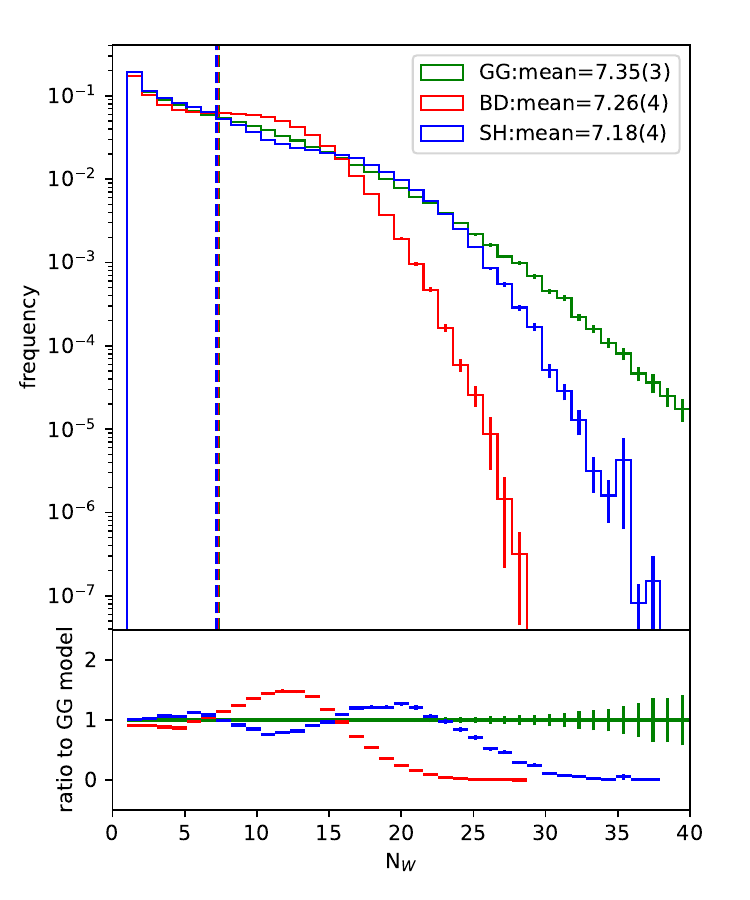}
    \caption{Distribution of number of wounded nucleons obtained from 50000 events using each one of the models from section \ref{sec:newMCG}. The impact parameter was allowed to vary between 0 and \unit[20.0]{fm}.}
    \label{fig:nw-bmax20}
\end{figure}

Using the Glauber model defined in the previous section, we study the distribution of number of wounded nucleons, \NW, in pPb events. Impact parameter is sampled linearly from $b_{min}=\unit[0]{fm}$ to $b_{max}=$\unit[20.0]{fm}. Fig.~\ref{fig:nw-bmax20} shows the \NW distribution for the different models from 50000 independent events.

As expected, the models accounting for cross section fluctuations (SH and GG) are able to reach further out in the tail of large \NW. Looking back at Fig.~\ref{fig:xs-vs-b-tune1}, we can see that the BD model reaches out until around \unit[1.4]{fm}, i.e., nucleons need to be at least at this distance to interact. As for the GG model, the minimum distance of interaction depends on the sampled cross section, however, it is at maximum about \unit[2.2]{fm}. On the other hand, the SH cross sections can extend until over \unit[3]{fm} and it still has a lower tail than the GG model. That happens because, when the GG model fluctuates to a larger spacial extension, all of the nucleons within that given radius will be wounded, whereas the SH model may miss nearby nucleons at the expense of gaining nucleons further away. Therefore, even with a larger extension it may not give such large \NW values as the GG model can, which shows that this tail of the \NW distribution is more sensitive to the largest cross section to which the model can fluctuate rather than to how far the interaction range for that model extends.

\begin{figure}[!ht]
    \centering
    \includegraphics[width=0.8\linewidth]{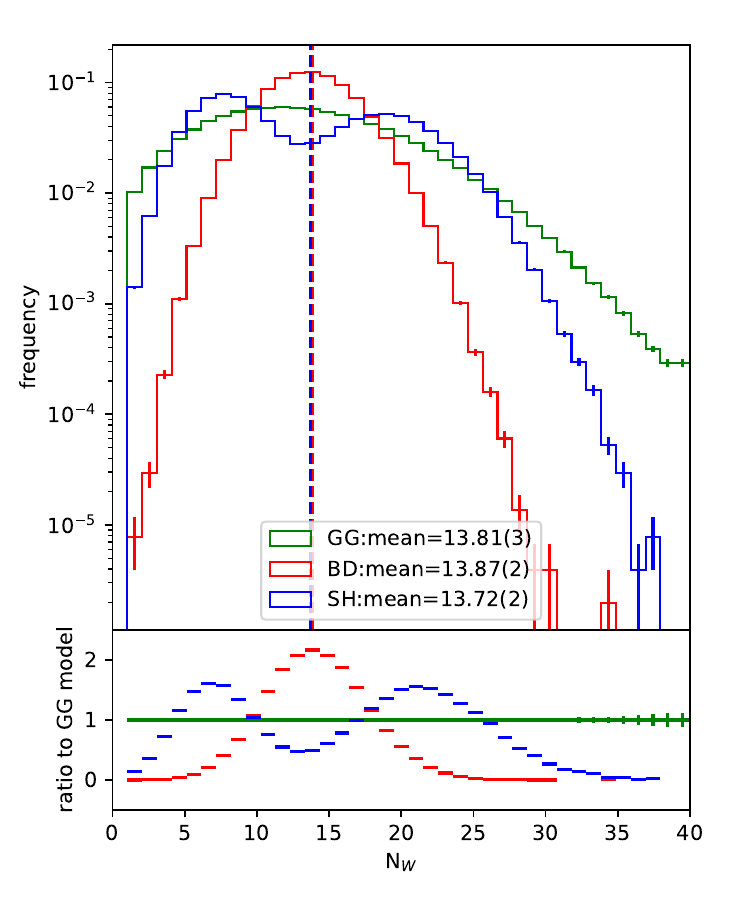}
    \caption{Distribution of number of wounded nucleons obtained from 50000 events using each one of the 3 studied models. The impact parameter is fixed at $b=$\unit[0.0]{fm}.}
    \label{fig:nw-b0}
\end{figure}

To illustrate this point, we fix the impact parameter to $b=0$. In that case, the \NW distribution obtained is the one in Fig.~\ref{fig:nw-b0}. In this case, the GG and SH models become even more different from each other. By fixing $b$ at the region of highest nucleon density, we wash away the effects of the fluctuations in nucleon positions. The result is that the \NW distribution from the BD model becomes much more peaked around the value it would provide in case there were no nucleon position fluctuations and the nucleus were simply a smooth density. The GG model also shows a more peaked distribution even though it still comes with the additional contribution from the fluctuating cross section. Interestingly, the SH model leads to a doubly peaked structure when we dial down the effect of nucleon fluctuations by fixing $b$, which is an effect from having the projectile nucleon fluctuate between two GW states, constraining the cross section for interaction between that nucleon and any of the target nucleons to one of two different values only, while the fluctuations of the target nucleons are averaged out. All of the models still provide very similar average numbers of wounded nucleons, reflecting the fact that they were all set up to provide the same average cross section.

Therefore, we see the very well established effect that fluctuations of the cross section give longer tails in the \NW distribution. We also see that they provide larger tails towards the small \NW direction, since the cross sections also fluctuate to small values. However, these are effects of the fluctuation of the integrated cross section. After all, for a given GG cross section, the only other fluctuation in play is the nucleon distribution. Apart from that, this model is completely deterministic at each given value for the cross section. On the other hand, if we consider the SH model, even if we fix the GW state configuration (and thus the integrated cross section) and nucleon positions, there is still a stochastic distribution of \NW. That is because, at each nucleon-nucleon distance, the impact parameter profile of the cross section determines only a probability for their interaction, which is not constrained to be either 1 or 0 but is continuously decreasing with distance between the two. This type of fluctuation also affects the \NW distribution, given the argument above, but, at least with only two GW states, it is not enough to reach the extremes of the distribution as far as the GG model can reach with only the integrated cross section fluctuations.

\begin{figure}[!ht]
    \centering
    \includegraphics[width=0.8\linewidth]{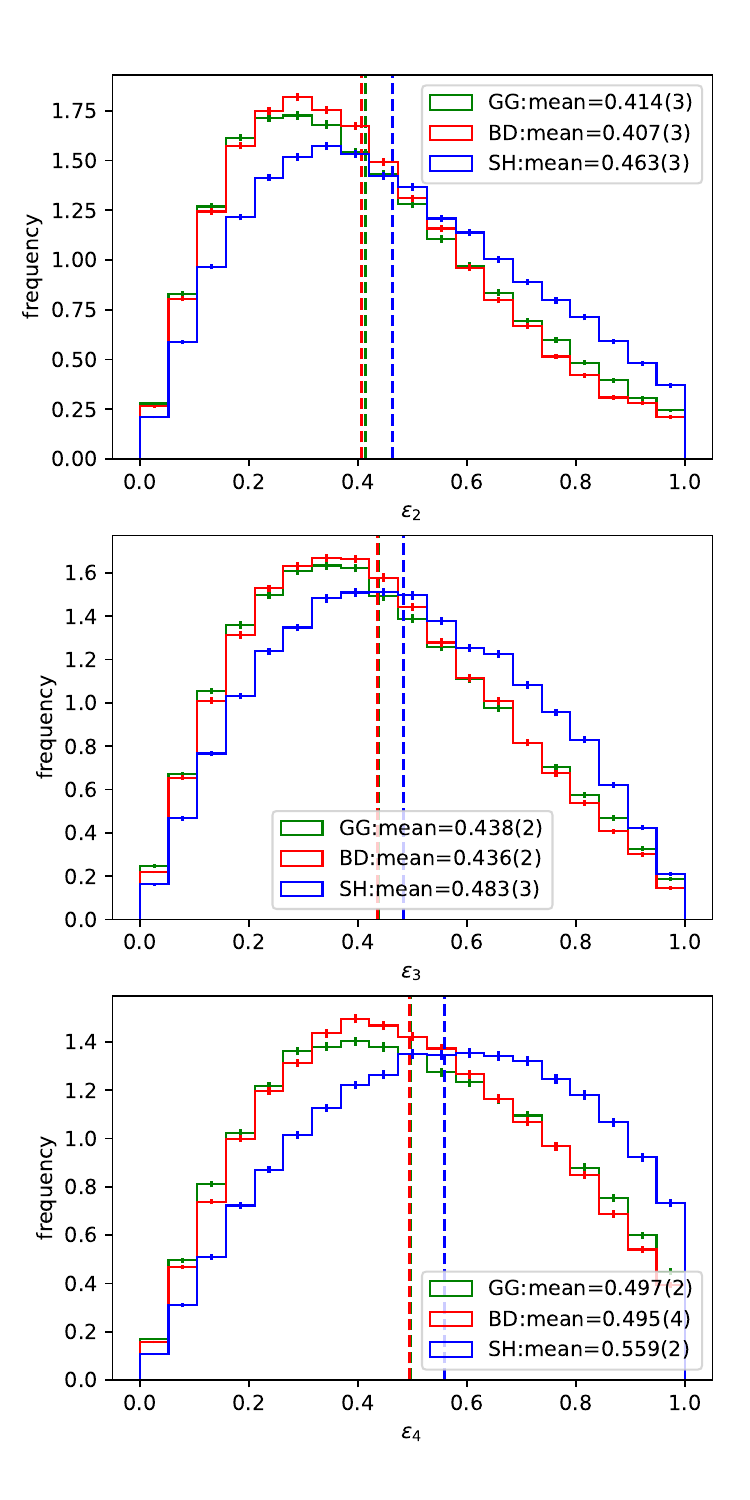}
    \caption{Distribution of the $\epsilon_n$  for (top) $n=2$, (middle) $n=3$ and (bottom) $n=4$ obtained in all the events with $b_{max}=$\unit[20.0]{fm} as calculated from equation \ref{eq:eccentricity}.}
    \label{fig:ecc-bmax20}
\end{figure}

Nonetheless, it should be expected that this additional fluctuation present in the SH model can also affect the geometry of the distribution of wounded nucleons. In a black-disk-like model, such as the GG one considered here, the cross sections do not provide any additional geometry fluctuation. The geometric aspect is reliant solely on the fluctuations of the nucleon positions. Therefore, it is interesting to look at the geometry of the wounded nucleon distributions as this is often used as an important input for hydrodynamic simulations for example. For that, we look at the n-th order eccentricities:

\begin{equation}
    \label{eq:eccentricity}
    \epsilon_n =\frac{ |\langle r^ne^{in\phi} \rangle|}{\langle r^n \rangle} ,
\end{equation}

\noindent where the averages are over the coordinates of the wounded nucleons, $r$ is the distance from the center of the wounded nucleon distribution and $\phi$ is the angle from the participant plane, computed as:

\begin{equation}
    \label{eq:planeangle}
    \Psi_n=\frac{1}{n}\arctan\left(\frac{\langle \sin(n\phi) \rangle}{\langle \cos(n\phi) \rangle}\right) + \frac{\pi}{2}\delta_{n2},
\end{equation}

\noindent where the $\pi/2$ shift is present in the 2-order plane as a convention\footnote{This way, the participant plane angle corresponds to the short axis of the ellipse instead of the long one.}. Fig~\ref{fig:ecc-bmax20} shows the distribution of the values of $\epsilon_n$  for $n \in [2,3,4]$ in those same events of Fig.~\ref{fig:nw-bmax20}, i.e., for $b_{max}=$\unit[20.0]{fm}.

\begin{figure}[!ht]
    \centering
    \includegraphics[width=0.8\linewidth]{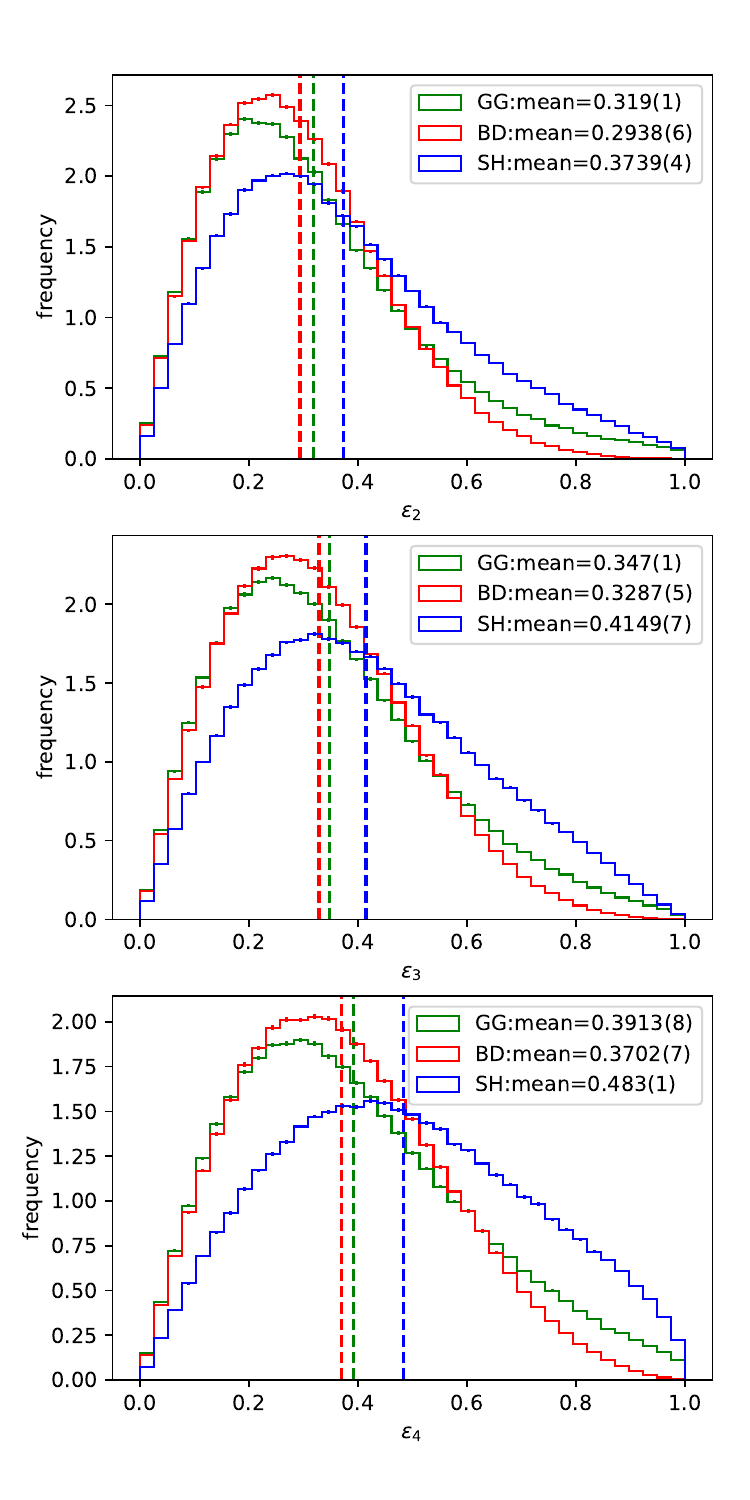}
    \caption{Distribution of the $\epsilon_n$  for (top) $n=2$, (middle) $n=3$ and (bottom) $n=4$ obtained in all the events with $b_{max}=$\unit[0.0]{fm} as calculated from equation \ref{eq:eccentricity}.}
    \label{fig:ecc-b0}
\end{figure}

One immediately notices that the two black-disk based models have very similar mean values and distributions despite their very different natures regarding fluctuations. On the other hand, the SH model is notably unlike them and tends to give substantially larger eccentricities. Again, we can fix $b=0$ in order to minimize the effects from the nucleon fluctuations and focus on the different cross section models. This way, we also remove the contribution from the slope in the nucleon density to the eccentricities. Fig.~\ref{fig:ecc-b0} shows these results.

In this case, the difference between the BD and GG models is more visible, although they are still very similar when compared to the SH model. In the BD model, the only way to create a non-zero eccentricity is via nucleon position fluctuations. As for the GG one, the only additional source of fluctuation is the integrated cross section and, therefore, the interaction range. Therefore, fluctuations in the integrated cross sections also contribute to fluctuations in the geometry of the wounded nucleon distribution and not only to the number of wounded nucleons, even though the contribution of the nucleon position fluctuations seems to be more important and partially washes away this effect. 

On the other hand, the SH model provides eccentricity distributions notably different from the other two models in both the inclusive events (Fig.~\ref{fig:ecc-bmax20}) and the fixed impact parameter ones (Fig.~\ref{fig:ecc-b0}). The additional fluctuations given by the non trivial impact parameter dependence clearly allow for broader $\epsilon_n$ distributions for all the values of $n$ calculated. The mean values obtained for all of the eccentricities calculated are also consistently larger than the ones from the other two models, with differences of up to about 20\%. This shows that the fluctuations in the cross sections given by a non trivial impact parameter dependence are extremely important when evaluating the geometry of the \NW distribution obtained with a certain Glauber model.

\section{Conclusions}
\label{sec:conclusions}

We investigated the effect of a non-trivial impact parameter dependence of the inelastic cross section on \NW and $\epsilon_n$ using the KMR/SHRiMPS model. One of the advantages of this model is that the impact parameter dependence comes out of the calculation alongside the actual cross section with no need for additional assumptions. It was shown that, even with only two GW states, the distribution of the number of wounded nucleons presents a large \NW tail, much larger than when compared to the BD model, and not too far from the GG one (Figs.~\ref{fig:nw-bmax20} and ~\ref{fig:nw-b0}). It was also shown that the non trivial impact parameter dependence of the cross sections in the SH model have a large impact in the geometry of the \NW distribution (Figs.~\ref{fig:ecc-bmax20} and ~\ref{fig:ecc-b0}).

Therefore, we see that integrated cross section fluctuations are essential for describing the shape of the \NW distribution but can also affect the geometry of the distribution (difference between the GG and BD models in Fig.~\ref{fig:ecc-b0}), although the nucleon fluctuations still dominate. On the other hand, the additional fluctuations obtained from non trivial impact parameter dependence are extremely important in defining the geometry of the \NW distribution, whereas they are not as significant in the definition of the shape of the \NW distribution as the value of the integrated cross sections. This is seen from the fact that the larger cross sections reached by the GG model lead to a larger \NW tail than the SH gives with a larger interaction range (Fig.~\ref{fig:nw-bmax20}), whilst the SH model predicts significantly large $\epsilon_n$ values.

Overall, this work has shown that, for a complete description of the initial state in pA collisions, both the integrated cross section fluctuations and the non-trivial impact parameter dependence need to be accounted for. This was done using a well motivated model for the cross sections, the KMR/SHRiMPS model, which includes both effects. Furthermore, this model can easily be extended for A+A collisions.

\section*{Acknowledgemet}

I would like to thank Korinna Zapp for the valuable discussions and for guiding me through the development of this work. I would also like to thank Frank Krauss and Korinna Zapp for the collaboration on the SHRiMPS model application for pA collisions. This study is part of a project that has received funding from the European Research Council (ERC) under the European Union's Horizon 2020 research and innovation programme  (Grant agreement No. 803183, collectiveQCD).

\bibliographystyle{unsrt} 
\bibliography{refs}

\end{document}